\documentstyle[12pt,twoside,fleqn,espcrc1]{article}
\title{NUCLEAR GAMMA RAY LINE ASTRONOMY 
IN THE PERSPECTIVE OF THE INTEGRAL SATELLITE}

\author{M. Cass\'e\address{Service d'Astrophysique, DSM/DAPNIA/CEA \\ 
 Orme des Merisiers 91191 Gif/Yvette France \\}%Ž
  , E. Vangioni-Flam\address{Institut d'Astrophysique de Paris, CNRS\\
98bis bd Arago 75014 Paris France}%
 and 
J. Paul\address{Service d'Astrophysique, DSM/DAPNIA/CEA\\
Orme des Merisiers 91191 Gif/Yvette France}}
\begin{document}
\maketitle
\begin{abstract}

We present a broad overview of the principal processes and astrophysical sites of gamma-ray line 
production and review the main pre-INTEGRAL satellite observations to set the stage to the next European era 
of gamma-ray line astronomy.
\end{abstract}

\section{INTRODUCTION}

Atoms in the universe have  three sources, nuclear fusion in two astrophysical contexts and 
nuclear break up or spallation. The primordial source is the big-bang with its associated nucleosynthesis. 
Operating at a temperature of about a billion degrees, it is responsible for the production of the lightest isotopes 
H, D, $^3$He, $^4$He, $^7$Li and lasts only about 3 minutes. Stars pursue the nuclear complexification of matter 
through a series of fusions at high temperature, building up progressively the composition of matter observed 
today in galaxies. In hydrogen  burning, H is transformed into He and in more advanced stages, all nuclei, 
between carbon and uranium, are synthesized by thermonuclear fusion at temperatures ranging from ~ 10$^7$ to 
5.10$^9$ K. 
A distinct process, of non thermal nature, puts the last touch to this nuclear evolution. 
Cosmic rays (through rapid p and alpha interactions with CNO in the interstellar medium, ISM) and  fast nuclei 
(rapid alpha, C and O fragmenting on interstellar H and He) produce $^6$Li, $^7$Li, $^9$Be, $^{10}$B, $^{11}$B.
To summarize very briefly, the nuclear complexification of matter has been laborious and 
slow:  at birth the galactic composition was close to that emerging from the big-bang (H = 0.76, He = 0.24, by 
mass) and presently, i.e. about 15 billions of years later, it contains a small proportion of "metals" (H = 0.70, 
He = 0.28, Z = 0.02, where Z is the "metallicity" comprising all species heavier than $^4$He).
The abundances measured in various astrophysical objects, however, reflect the cumulated 
nucleosynthesis in the whole past. The great merit of gamma-ray line astronomy is to reveal the present nuclear 
activity in the Galaxy. However, the number of measurable gamma-ray emitters is restricted compared to the 
whole list of nuclei of the periodic table. Only a handful of isotopes is available: $^7$Be, $^{22}$Na, $^{26}$Al, $^{44}$Ti, 
$^{56}$Co, $^{57}$Co, $^{60}$Fe. Those have mean lifetimes between 0.3 to 10$^6$ years and are produced with significant 
abundances by stars (table 1). 
Gamma ray line astronomy has a unique potential to provide information on nuclear processes and 
high energy interactions going on presently in the Universe. Moreover, space is transparent to gamma-ray lines 
up to a redshift of 100 and the fluxes are almost not affected by their travel in our Galaxy. The informations 
carried by gamma ray lines - intensity, line profile and width - are translated respectively in physical terms 
through astrophysical models of the sources. The intensity of a given source at a given energy leads to the 
abundance of the emitting isotope and to physical conditions prevailing to its formation (temperature, density or 
flux of energetic particles if the formation process is non thermal). The line shift compared to the laboratory 
position is used to derive, through the Doppler effect, the expansion velocity of the emitting material (wind 
velocity, supernova and nova envelope expansion) and/or the gravitational shift suffered by the radiation if it is 
released close to a compact object. The line width reflects the temperature of the medium, if thermal, or the 
energy spectrum of the fast particles, if non thermal.

\section{GAMMA-RAY LINE EMISSION REQUIREMENTS} 

How to populate excited levels of nuclei? In principle, two means are available, impliying 
respectively radioactive or stable nuclei.
The first way rely on gamma ray emission in the course of radioactive decay. The emission is in 
this case delayed by the radioactive lifetime. This process is only efficient if some (rather) abundant nuclei made 
under the form of their radioactive progenitor. The best example is $^{56}$Ni decay (table 1).
By its very nature thermal nucleosynthesis produces proton-rich isotopes at high temperatures and 
densities in star interiors, opaque to gamma rays. Thus radioactive species have to be transported in the ISM 
before decaying to deliver their gamma signature. Various means are used by nature to sprinkle the ISM with 
live radioactive nuclei, including stellar winds and explosions, in short, dynamical events. 
The second way to produce gamma ray lines is nuclear excitation of stable and abundant  isotopes 
by fast (5-50 MeV/n) particles. Gamma emission is prompt due to the very short lifetime of the excited nuclear 
levels. These non-thermal processes, in order to be observable, require high fluxes of low energy particles. We 
do not consider the annihilation line emission at 511 keV since this subject deserves by itself a full 
developement.

\section{ASTROPHYSICAL BACKGROUND}

 Thermal nucleosynthesis of radioactive isotopes is present under two forms; i) hydrostatic 
nucleosynthesis, proceeding in Wolf-Rayet stars (massive star supporting heavy mass loss through intense 
winds) and AGB stars (red giants developing thermal pulses); ii) explosive nucleosynthesis, occuring in core 
collapse supernovae (SNII, SNIb, c), thermonuclear supernovae in binary systems (SNIa) and novae. Non 
thermal processes take place in the ISM bombarded by fast projectiles and in solar flares. 

\subsection{Thermal nucleosynthesis}
a. Hydrostatic nucleosynthesis

 This nucleosynthesis is active in rather quiet stellar stages, where the temperature and the 
density are constant over a long period of time. As far as gamma-ray line astronomy is concerned, the principal 
process of interest is the production and ejection of $^{26}$Al by WR stars, that seem to be the best canditates 
according to the recent analysis of the observations made by the COMPTEL  experiment on board of the CGRO 
satellite.
During H burning, 26Al is produced via the radiative proton capture by 25Mg. The intense wind remove 
the unprocessed envelope exposing the convective core, and fresh $^{26}$Al is carried away by the wind before 
decaying. The decay, finally, produces a line at 1.809 MeV in the transparent ISM. The mass of $^{26}$Al generated  
is estimated typically to about 10$^{-4}$ Mo per WR. AGB stars could also eject $^{26}$Al but the astrophysical scenario 
is not precisely known.

b. Explosive nucleosynthesis

The time scale of explosive nucleosynthesis is so short that beta decay has no time to operate. 
Under these conditions, a host of isotopes are made under the form of their radioactive (proton-rich) progenitor. 
Among them, $^{56}$Ni, $^{57}$Ni and $^{44}$Ti are of the highest interest for gamma-ray line astronomy. $^{26}$Al is 
presumably also formed in core collapse supernovae and additionally by neutrino spallation during the 
explosion.
The mass of the radioactive isotopes generated in core collapse supernovae  depends on the temperature and 
density reached behind the shock wave and on mass cut (frontier between the imploding core giving rise to a 
neutron star and the ejected material). These parameters are uncertain since they are sensitive to fine details of 
modelization of the  presupernova structure and evolution and on the detailed hydrodynamical treatment of the 
explosion itself. Moreover the escape of gamma's depends on internal mixing and unstabilities.
In thermonuclear supernovae (SNIa), the scenario is different: the exploding object is a white dwarf 
(WD) overloaded by the material accreted from a companion. When the critical (Chandrasekhar) mass of about 
1.4 Mo is exceeded, the star looses its stability.  Degenerate Carbon burning becomes  explosive and a large 
fraction of the WD is transformed into $^{56}$Ni. 
Nova explosions, originating also from WD 's in binary systems, are much more frequent than 
SN ones, but they are less spectacular, and their gamma-ray line emission is only observable in the galaxy. 
Accretion proceeds to a different rate than that leading to SNIa explosions. The explosive eruption of these 
objects should release substantial amounts of $^7$Be, $^{22}$Na and $^{26}$Al. Here the main uncertainties concern the 
amount of matter ejected and the treatment of convection which is always a problem in the modelization of 
stellar objects.

\subsection{Non thermal nuclear excitation}

The second mode of production of gamma-ray lines is associated with flows of fast 
particles. The interaction of the energetic nuclei generated by shock wave acceleration with the ambient 
target medium could, in principle, produce a wealth of gamma ray lines. Observations of C and O lines at 
4.4 and 6.1 MeV would be the clue to the irradiation of molecular cloud by nuclei of helium, carbon and 
oxygen produced and accelerated to moderate energy by massive stars (WR, SN).

\section{OBSERVATIONAL HIGHLIGHTS}

\subsection{Galactic Disk}

The 1.809 MeV emission due to $^{26}$Al decay deduced from the COMPTEL observations through a 
sophisticated procedure has been  correlated with the free-free emission of the galactic disk observed by COBE in 
the microwave regime. Extended sources as Vela (SNR, WR) and Cygnus (Massive Stars) have been located. 
The distribution of $^{26}$Al emission seems to indicate that massive stars are the main sources of $^{26}$Al.

\subsection{SN 1987A}

This core collapse supernova has exploded in the Large Magellanic Cloud, and it appeared in the 
southern sky in Februrary, 1987. It has been observed by many telescopes including neutrino ones. Its light 
curve has shown an exponential decay of 77 days, commensurate with the mean lifetime of $^{56}$Co. The passage 
from opacity (inherent to stars) to interstellar transparence has been induced by the thinning of the remnant 
under the effect of its expansion. Moreover gamma-ray lines at 0.847 and 1.238 MeV from $^{56}$Co decay have 
been directly detected and also the 0.122 MeV line from $^{57}$Co decay. Gamma-ray  emission has taken place 
earlier than predicted, leading to speculate on internal mixing. The 57/56 isotopic ratio deduced is close to solar. 
The amount of iron synthesized is 0.07 Mo. This is the most precise determination of iron production by a 
supernova never achieved. These observations can be considered as a bright confirmation of the theory of 
explosive nucleosynthesis.

\subsection{Cas A}

Cas A is a young SN remnant (~ 300 yr)  located at a distance of 2.8 kpc. The gamma-ray line from $^{44}$Ti decay 
has been detected by COMPTEL. Thanks to a specific supernova model, the mass mass of $^{56}$Ni produced and 
ejected has been deduced from the mass of $^{44}$Ti observed. The $^{56}$Ni mass is so high that  this supernova should 
have been easily visible at the time of its maximum brightness, whereas no historical record bears its trace. The 
Cas A puzzle can be solved, however, if one imagine that the explosion of the central object has been hidden by 
the thick and dusty wind of the progenitor star.

\subsection{Orion and Vela complexes}
The COMPTEL satellite has perhaps detected a flux of gamma rays characteristic of the 
deexcitation of carbon and oxygen from the Orion  molecular cloud complex, located at a distance of about 500 
light-years from our Solar System. A recent spectrum taken from the Vela superbubble at about the same 
distance is similar to that of Orion. In these regions, marked by the explosion of several supernovae, massive 
stars abound. However the detections are still marginal and debated. Future observations will clarify the 
situation.

\section{INTEGRAL MISSION}

The ESA (European Space Agency) scientific mission INTEGRAL (International Gamma-Ray 
Astrophysics Laboratory) is dedicated to the fine spectroscopy, owing to the use of germanium detectors $(E/\Delta E 
= 500)$ and imaging of celestial gamma-ray sources in the energy range 15 keV to 8 MeV. INTEGRAL will be 
launched in the beginning of the next decade by a Russian PROTON rocket into a highly eccentric 72-hour 
orbit. The nominal lifetime of the observatory will be 2 years with possible extension to up to 5 years. Most of 
the observing time will be made available to the worldwide scientific community. The scientific goals of 
INTEGRAL are addressed through the simultaneous use of high resolution spectroscopy with fine imaging and 
accurate positioning of celestial sources in the gamma-ray domain. Fine spectroscopy over the entire energy 
range will permit spectral features to be uniquely identified and line profiles to be determined for physical studies 
of the source region.
The gamma-ray emission from the galactic plane will be mapped on a wide range of angular 
scales from arc-minutes to degrees in both discrete thermal and nonthermal nucleosynthesis lines, from $^{26}$Al, 
C* and O*  together with the 511 keV, and the wide band continuum. At the same time, source positioning at 
the arc-minute level within a wide field of view, of both continuum and discrete line emissions, is required to 
allow an extensive range of astrophysical investigations to be performed on a wide variety of sources, both 
targeted and serendipitious, with a good chance of identification at other wavelengths.
 Measurements with INTEGRAL of the shapes of the gamma-ray line profiles from supernovae, 
particularly SNIa in the Virgo cluster of galaxies, will provide information about the expansion velocity and 
density distribution inside the ejected envelope, whilst the relative intensities of the lines will provide direct 
insight into the physical conditions at the time of the production.

\section{CONCLUSION}

A golden age of nuclear gamma ray line astronomy is opening up in Europ, at the point of 
convergence of nuclear physics and astrophysics. Time is ripe to join our efforts. 
Explicit references can be found in the proceedings of the INTEGRAL symposia (Saint Malo, 
1997 (1) and Taormina, 1999 (2) ) published by the European Space Agency.

This work was supported in part by the PICS 319, CNRS at the Institut d'Astrophysique de Paris.

\end{document}